\documentclass[a4paper,twocolumn,english,pre,nofootinbib]{revtex4}
\usepackage[T1]{fontenc}
\usepackage[latin9]{inputenc}
\usepackage{color}
\usepackage{babel}
\usepackage{mathtools}
\usepackage{amsmath}
\usepackage{amssymb}
\usepackage{graphicx}
\usepackage[unicode=false,pdfusetitle,
 bookmarks=false,bookmarksnumbered=false,bookmarksopen=false,
 breaklinks=false,pdfborder={0 0 1},backref=false,colorlinks=true]
 {hyperref}
\graphicspath{{./Figures/}}

\begin{document}


\title{Fast rare events in exit times distributions of jump processes}

\author{Alessandro Vezzani}
\affiliation{Istituto dei Materiali per l'Elettronica ed il Magnetismo (IMEM-CNR), Parco Area delle Scienze, 37/A-43124 Parma, Italy}
\affiliation{Dipartimento di Scienze Matematiche, Fisiche e Informatiche,
Universit\`a degli Studi di Parma, Parco Area delle Scienze, 7/A 43124 Parma, Italy}
\affiliation{INFN, Gruppo Collegato di Parma, Parco Area delle Scienze 7/A, 43124 Parma, Italy}

\author{Raffaella Burioni}
\affiliation{Dipartimento di Scienze Matematiche, Fisiche e Informatiche,
Universit\`a degli Studi di Parma, Parco Area delle Scienze, 7/A 43124 Parma, Italy}
\affiliation{INFN, Gruppo Collegato di Parma, Parco Area delle Scienze 7/A, 43124 Parma, Italy}

\begin{abstract}
Rare events in the first-passage distributions of jump processes are capable of triggering anomalous reactions or series of events. Estimating their probability is particularly important when the jump probabilities have broad-tailed distributions, and rare events are therefore not so rare. We formulate a general approach for estimating the contribution of fast rare events to the exit probabilities in the presence of fat tailed distributions. Using this approach, we study three jump processes that are used to model a wide class of phenomena ranging from biology to transport in disordered systems, ecology and finance: discrete time random-walks, L\'evy walks and the L\'evy-Lorentz gas. We determine the exact form of the scaling function for the probability distribution of fast rare events, in which the jump process exits from an interval in a very short time at a large distance opposite to the starting point. In particular, we show that events occurring on time scales orders of magnitude smaller than the typical time scale of the process can make a significant contribution to the exit probability. Our results are confirmed by extensive numerical simulations.

\end{abstract}

\maketitle


A stochastic process that reaches a certain threshold value for the first time can trigger many events: a chemical reaction occurs \cite{Gardiner},  a target is reached \cite{Benichou05,TargetPRE10,Benichou-searchRMP,Godec16,Eliquantum18,Meyer21}, the dynamics of a financial process starts \cite{ChicheBouch14}, biological and ecological processes take place \cite{Pulkkinen13,Arani21}. The study of these triggering events and the probability of their occurrence is based on the knowledge of first passage probabilities \cite{Redner}.  As history-dependent quantities, first-passage probabilities are difficult to determine: general results are typically available for the average first-passage time and its lower-order moments \cite{Redner,ben-voit-PRep}, but knowledge of complete first-passage distributions is generally limited. Given the 'trigger' property of exit times, the tails of these distributions are particularly important as they allow the probability of rare anomalous events to be estimated, such as, for example, an exit in a very short time \cite{Lawley20}. These estimates are particularly relevant in the case of broad-tailed distributions, where rare events are not so rare. 

In particular, a few results are available for first-passage probabilities of jump processes, which are stochastic processes that involve random jumps between different states or positions, occurring at random times and with random magnitudes \cite{VanKampen}. Jump processes are so important in modelling the dynamics of stochastic processes in many fields  that even their one-dimensional formulation is of great relevance \cite{Chechkin05,Dudko08,Marconi08}. In the one-dimensional case, exit time probabilities refer to the process leaving a particular state or interval within a specified time. In stochastic processes with two alternative outcomes, the exit side can also be of importance \cite{Arani21}, for example to quantify observables such as transmission or backscattering probabilities \cite{Rotter}. 
A well know examples of jump processes is that of random-walks {(RWs)}, in discrete and continuous time \cite{weiss,RW_Klafter}, for which recent results have been obtained \cite{Levernier21,Klinger,exit}.  An interesting question is to determine the exit probabilities from the side of a domain opposite to the starting point. {A rare event in this case corresponds to a fast walker leaving the interval in a very short time, i.e. in a time where the typical distances covered are still much shorter than the size of the domain.}
To exit in a short time, the trajectory of the walker should correspond to a long jump, allowing it to travel far from the starting point. When fat-tailed distributions for jumps are present in the stochastic process, these very fast events {may occur} in timescales orders of magnitude shorter than expected, proving crucial in predicting anomalous behaviors.



Recently, we have investigated the role played in rare events by the so called big jump principle \cite{BJ1}. The principle explains extreme events in a wide class of systems with heavy tailed distributions not in terms of an accumulation of many small subevents but as an effect of only the biggest event, the big jump \cite{BJFoss}. The principle has been successfully applied to characterize the tail of the probability distribution, at distances much larger than their scaling length, for a wide class of jump processes for RWs involving L\'evy and sub-exponential statistics for space and time \cite{BJ1,BJ_sub_exp,BJ2,BJ3}, even in disordered settings such as the L\'evy-Lorentz gas \cite{LevyLorentzBarkai,LevyLorentz1}. 


In this letter, {we first extend the big jump approach to first passage probabilities and formulate the principle in a general way that makes it applicable to a large set of processes with jumps following a sub-exponential statistics, in the sense of \cite{BJFoss,Goldie}. In particular, for a walker starting at the origin, we analytically determine the probability of reaching for the first time a distance $X$ without being absorbed at the origin, in the asymptotic limit where $X$ is much larger than the scaling length of the process (i.e. in the limit of very fast atypical events). We consider three well known models of one-dimensional jump processes: the discrete time RW, the most basic jump process \cite{RW_Klafter}, the L\'evy walk, widely used to model animal movement and search patterns \cite{foraging1,foraging2}, and the L\'evy-Lorentz gas, a relevant model in the study of transport in complex and disordered media \cite{Wiersma2008}. For this last model, there is currently no known first-passage probability estimate \cite{Burov21,Bousige21}. Taking advantage of the new formulation,  we estimate  the exit probability from one side of a domain and show that, interestingly, fast rare events can make a significant contribution to the exit time probability even when the scaling length is diffusive and the bulk of the distribution is Gaussian. We test our predictions by comparing them with extensive numerical simulations, with very good agreement.} Our results show that the big jump principle can also be applied to first-passage problems, whereas previously it has only been used for probability densities. This broad applicability is based on the fact that the principle provides insight into the physical process that leads the walker to cover very large distances.


{\it The big jump for exit time probabilities}. 
{In a stochastic jump process, 
we call $P_X(T)$ the probability density of reaching a distance $X$ from the origin for the first time at time $T$ without being absorbed by the starting point, $x=0$. 
The characteristic length $\ell(T)\sim T^\gamma$ is defined by the scaling form of  $P_X(T)$ such that, for $X\sim \ell(T)$,  $P_X(T)\sim f(X/\ell(T))$ \cite{BJ1}.}   
If there are independent jumps (renewals) in the process, drawn from a sub-exponential distribution \cite{BJFoss,Goldie}, that allow to reach a distance $X>>\ell(T)$, then the big jump principle is expected to hold \cite{BJ1}.
In particular, the principle states that a process reaching a distance $X>>\ell(T)$ is triggered by a single very large jump, drawn from the subexponential distribution, while shorter distances of the order of $\ell(T)$ are reached by processes involving multiple jumps. This physical insight suggests that we can focus on the big jump and for $X>>\ell(T)$ write the probability density function (PDF) $P_X(T)$ as:
\begin{equation}
    P_X(T) \sim R(T_w)\cdot S(T_w)\cdot p_s(X).
    \label{eq:BJ}
\end{equation}
Here, $p_s(X)$ is the probability that the walker will reach a distance greater than $X$ in a single jump; $T_w$ is the time at which the big jump occurs, i.e. $T_w=T-\tau(X)$ 
where $\tau(X)$ is the time at which the walker reaches the distance $X$ in a single jump; the survival probability $S(T_w)$ is the probability that the walker is not absorbed before time $T_w$; and finally, $R(T_w)$ is the rate at which jumps are made at time $T_w$.  The $p_s(X)$ and $\tau(X)$ can be calculated directly by knowing the dynamics of the single jump, while the rate $R(T_w)$ and the survival probability $S(T_w)$ depend on the overall stochastic process. This factorization allows us to separate the calculation into quantities that depend on different characteristics of the stochastic process. {We note that the renewals, which allow the application of the big jump, may not coincide with all stochastic events in the dynamics (see, for example, the difference between scattering events and renewals in the L\'evy Lorentz gas).}

\begin{figure}
\includegraphics[scale = 0.62]{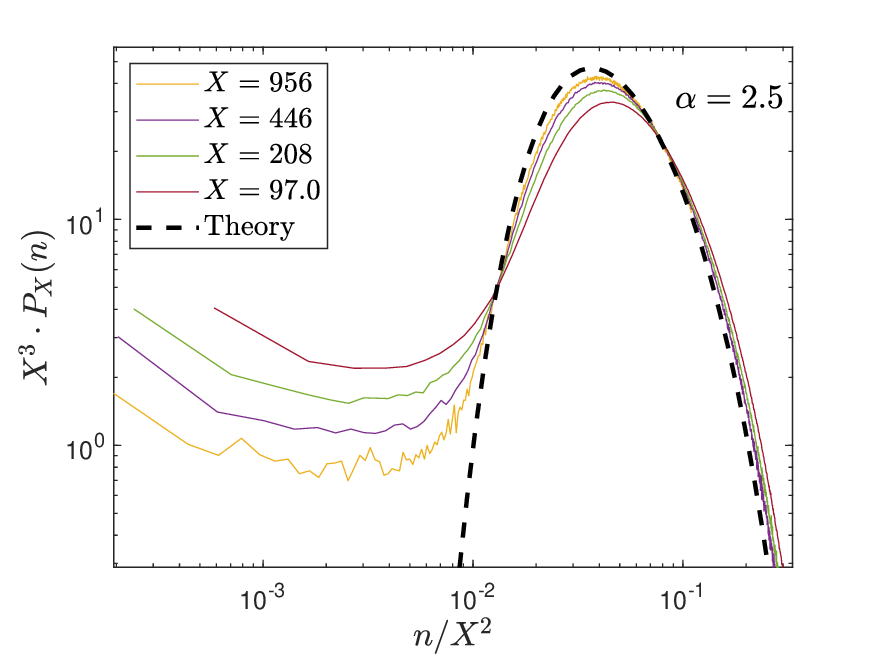}
\caption{The exit probability $P_X(n)$ for a L\'evy flight is plotted as a function of the rescaled exit time  $n/X^2$ ($\alpha>2$). We use a step length distribution $p(r)$ with a cut-off at short distances $r_0=1$. The typical dynamics is diffusive and, at large number of steps $n$, converges slowly to the theoretical prediction with no free parameters (Eqs (1,5,6) in \cite{exit}) which also provides the scaling factor $X^3$.  On the left,  very fast events are present with significant probability.}
\label{fig:flight_cenral}
\end{figure}

{\it The discrete time random walk}. In discrete time RWs \cite{RW_Klafter}, at each discrete time step the walker moves with probability $1/2$ to the right or to the left, with the step length $r$ drawn from the PDF $p(r)$. A finite second moment of $p(r)$ implies standard diffusion while, if the second moment diverges, the walker performs a symmetric L\'evy flight. We consider a $p(r)$ with a power law decay at large $r$:
\begin{equation}
    p(r) \sim \frac{\alpha r_0^\alpha }{r^{1+\alpha}} 
\label{eq:p}
\end{equation}
emphasizing that the principle holds for any sub-exponential $p(r)$ \cite{BJ_sub_exp}.
Here time is defined by the number of steps $n$ and $x(n)$ is the position of the walker after $n$ steps ($x(0)=0$). 
This process features standard diffusion with a scaling length $\ell(n)\sim n^{1/2}$ for $\alpha>2$ and a L\'evy superdiffusive dynamics with $\ell(n)\sim n^{1/\alpha}$, for $0<\alpha<2$. We focus on the exit time probability $P_X(n)$, i.e. the probability that the walker at step $n$ reaches a distance greater than $X$ for the first time without being absorbed at the origin. The numerical results in Figure \ref{fig:flight_cenral} show $P_X(n)$ as a function of the exit time  $n/X^2$, with the time rescaling determined by the growth of $\ell(n)$ (for $\alpha=2.5$, $X\sim\ell(n)\sim n^{1/2}$). The scaling form of $P_X(n)$ and the analytical plot in the dashed line were obtained in \cite{exit} using a continuous limit. Very fast events, occurring on much shorter timescales (i.e. for $n^{1/2}\ll  X $) are also observed with non-negligible probability. These are the rare events that we want to describe.

\begin{figure}
\includegraphics[scale = 0.62]{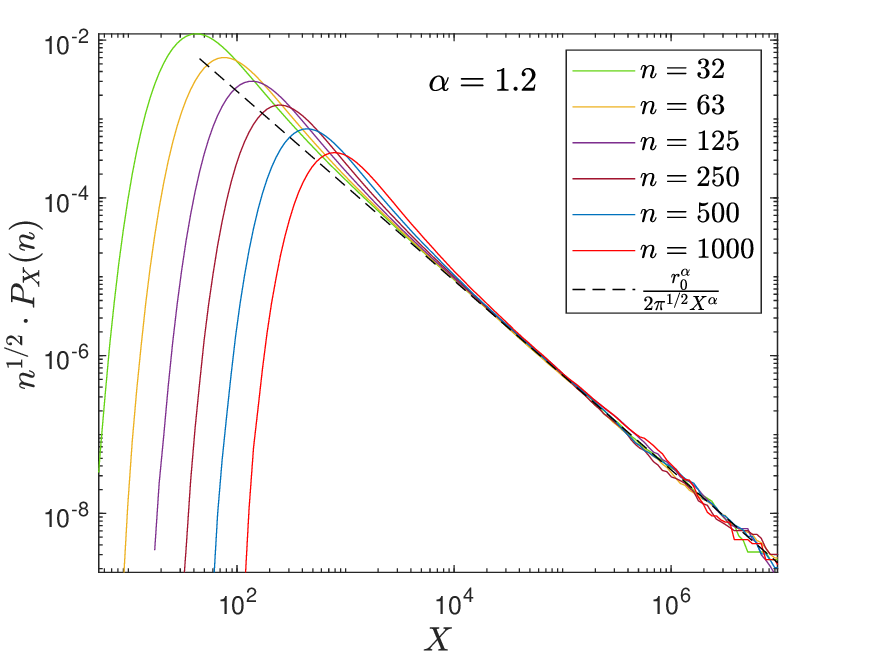}
\includegraphics[scale = 0.62]{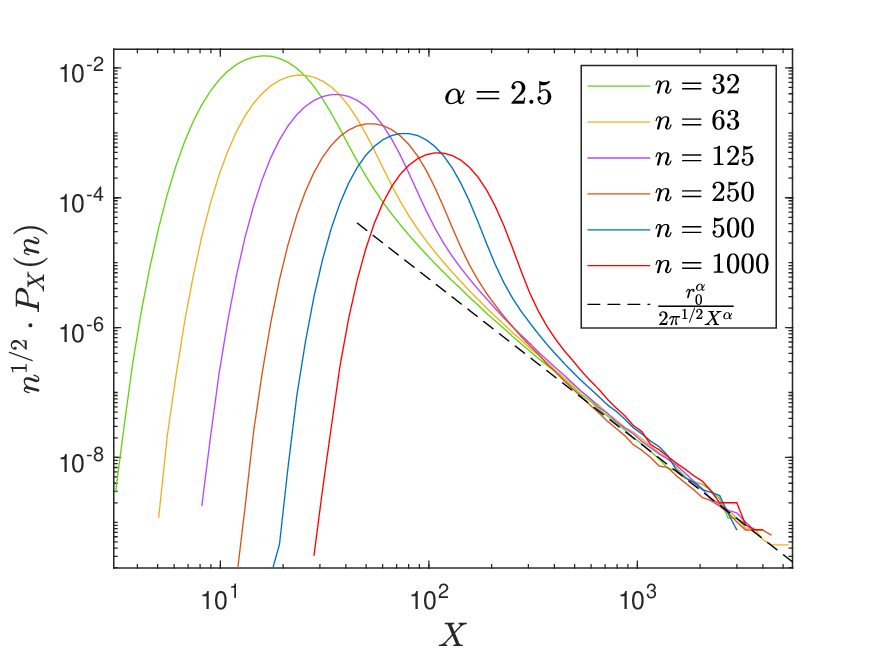}
\caption{Exit probability $P_X(n)$ for a L\'evy flight. $P_X(n)$ is multiplied by $n^{1/2}$ and plotted as a function of the distance $X$. Upper panel $\alpha=1.2$; lower panel $\alpha=2.5$. The dashed line shows the asymptotic analytical prediction in Eq. \eqref{eq:flight}.}
\label{fig:flight}
\end{figure}

The big jump principle states that the RW exits at time $n$ at a distance $X\gg \ell(n)$ if it is not absorbed at the origin, and if exactly at step $n$ the walker makes a jump in the positive direction, of length greater than $X$. The distance covered in the previous steps can be neglected. Now, in discrete time RW, jumps are instantaneous, so $\tau(X)=0$ and $T_w=T=n$. Moreover, $S(T_w)=S(n)$ can be obtained by the Sparre-Andersen theorem \cite{SparreAndersen,SparreAndersen2,SparreAndersen2_Maj}, which states that, for a general jump processes in a semi-infinite line, the survival probability after $n$ steps decays asymptotically as $S(n)\sim (\pi n) ^{-1/2}$. The probability of making a jump greater than $X$ is $p_s(X)= \frac{1}{2} \int_X^\infty dr p(r)$ where the factor $1/2$ accounts for the $1/2$ probability of forward jumps. The renewal rate is trivially one, so for $n\gg 1$ and $X\gg \ell(n)$ Eq. \eqref{eq:BJ} reads:
\begin{equation}
    P_X(n)\sim S(n) \frac{1}{2}
    \int_X^\infty dr p(r) \sim
    \frac{1}{2 (\pi n)^{1/2}}\cdot \frac{r_0^\alpha}{X^\alpha}.
\label{eq:flight}
\end{equation}
Eq. \eqref{eq:flight} is compared with numerical simulations in Figure \ref{fig:flight}, showing an excellent agreement in the asymptotic regime at large $X$, both for the L\'evy flights $\alpha<2$ and for normal diffusion $\alpha>2$. On the other hand, the peaks of the distributions at shorter $X$ occur when the system size is comparable to the scaling length  $\ell(n)$, according to the analytical predictions in \cite{exit}, as shown in Figure \ref{fig:flight_cenral} for $\alpha>2$.

{\it The L\'evy walk}. The L\'evy walk \cite{RW_Klafter,LevyWalkRv} is a continuous-time process in which a walker takes random steps of duration $t$ at constant velocity $v$. In one dimension, each step is covered with equal probability in the positive and the negative direction, and the duration of the steps follows the PDF $p(t)$. {We consider systems in which $p(t)$ at large $t$ is again a power law:
\begin{equation}
    p(t) \sim
\frac{\alpha t_0^\alpha}{t^{1+\alpha}}.
\label{eq:p_t}
\end{equation}
The scaling length of the stochastic process is $\ell(t)\sim t^{\gamma}$, with $\gamma=1$ for $\alpha<1$, $\gamma=1/\alpha$ for $1<\alpha<2$  and $\gamma=1/2$ for $\alpha>2$ \cite{LevyWalkRv}. In addition, the single step contains another scaling length growing linearly with time, due to the constant velocity motion. Therefore, for $\alpha<1$ in a single step the walker cannot cover a distance much larger than $\ell(t)$ and the single big jump approach cannot be applied. On the other hand, for $\alpha>1$ as well as for other sub-exponential PDFs (e.g. Weibull), the principle can be used efficiently \cite{BJ1,BJ_sub_exp}. 

In L\'evy walks the motion within a step is ballistic, so $\tau(X)=X/v$ and $T_w=T-X/v$. The number of steps in a time $T_w$ is $n=T_w/\langle t \rangle$, where $\langle t \rangle=\int dt' t' p(t')$ is the average duration of the steps. The Sparre Andersen theorem can again be used and gives {$S(T_w) \sim (\pi T_w/\langle t \rangle)^{-1/2}$}, and the jump rate is constant, $R(T_w) = 1/\langle t \rangle$ \cite{BJ1,BJ_sub_exp}.
Finally, $P_s(X)= \frac{1}{2} \int_{X/v}^\infty dt p(t)$ where again we take into account of the ballistic motion within a step and of the $1/2$ probability of forward jumps.
Summing up from Eq. \eqref{eq:BJ} we obtain: 
\begin{equation}
    P_X(T) 
    \sim \frac{1}{(vT)^{\alpha+1/2}}
    \frac{\langle t \rangle^{-1/2} t_0^\alpha }{2 (\pi (1-y))^{1/2}y^\alpha}
    \label{eq:walk}
\end{equation}
where $y=X/(vT)<1$. The ballistic motion in the big jump naturally introduces a new length scale, $vT$, which is the maximum distance that can be travelled in a time $T$.
Notice that $\langle t \rangle$ in Eq. \eqref{eq:walk} depends on the whole shape of $p(t)$ and not on its far tail only. In the simulations we consider a distribution with a cut-off at $t=t_0=1$. In Figure \ref{fig:walk} we show that $P_X(T)$ plotted against $X/(vT)$ indeed converges to Eq. \eqref{eq:walk} at large times and distances, showing that the big jump estimate is correct.

\begin{figure}
\includegraphics[scale = 0.62]{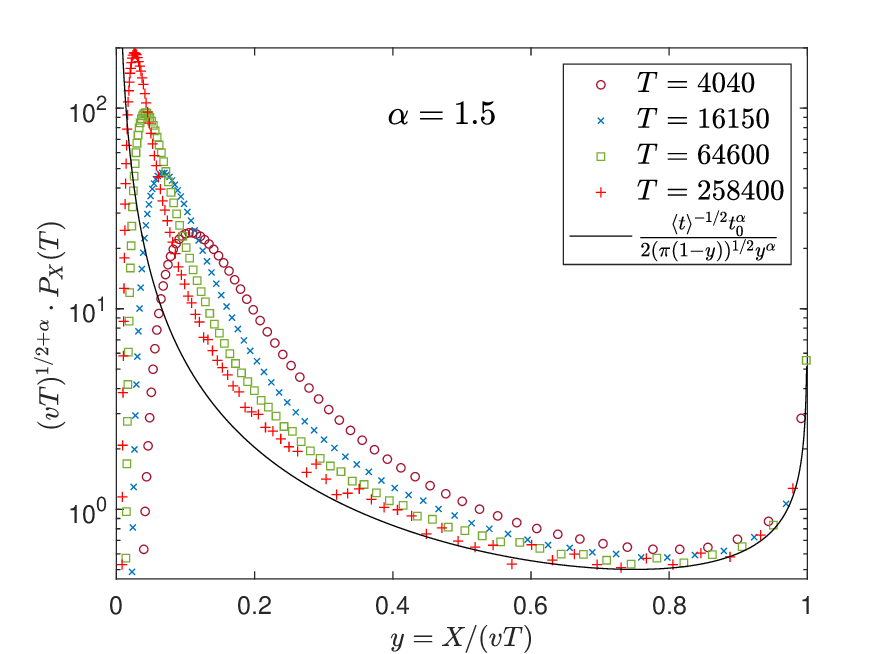}
\caption{Exit probability $P_X(T)$ for a {L\'evy walk}. $P_X(T)$ is multiplied by $(vT)^{1/2+\alpha}$ and plotted as a function of the rescaled distance $X/(vT)$, $\alpha=1.5$. The continuous line shows the asymptotic analytical prediction in Eq. \eqref{eq:walk}. }
\label{fig:walk}
\end{figure}

\it The L\'evy Lorentz gas}. The L\'evy Lorentz gas \cite{LevyLorentzBarkai,LevyLorentz1} is a system of scatterers randomly placed in one dimension. The distances between scatterers are drawn again from a power law PDF $p(l)$ {at large $l$:
\begin{equation}
    p(l)\sim
\frac{\alpha l_0^\alpha }{l^{1+\alpha}}.
\label{eq:p_l}
\end{equation}
A RW is naturally defined on the L\'evy Lorentz gas:  the walker moves at constant speed $v$ and is reflected with probability $\epsilon$ ($0<\epsilon<1$) when it hits a scatterer \cite{LevyLorentz2Univ}. We focus on the case where the walker starts at $t=0$ in a scattering site placed at $x=0$ \cite{LevyLorentzBarkai,LevyLorentz1}. 
For this model, the PDF of the walker position has recently been studied \cite{LevyLorentz1,LevyLorentz2Univ,Artuso_2018,Bianchi2016,Bianchi2020}: the scaling length reads $\ell(T)\sim T^{\frac{1}{1+\tilde \alpha}}$, 
where $\tilde \alpha=\alpha$ if $\alpha<1$ and  $\tilde \alpha=1$ if $\alpha>1$
\cite{LevyLorentz1} and the behaviour of the PDF at large distances has been estimated using the big jump principle \cite{LevyLorentz1,BJ1,zamparo}. 
However, no results are yet known for the exit time probabilities.

\begin{figure}
\includegraphics[scale = 0.62]{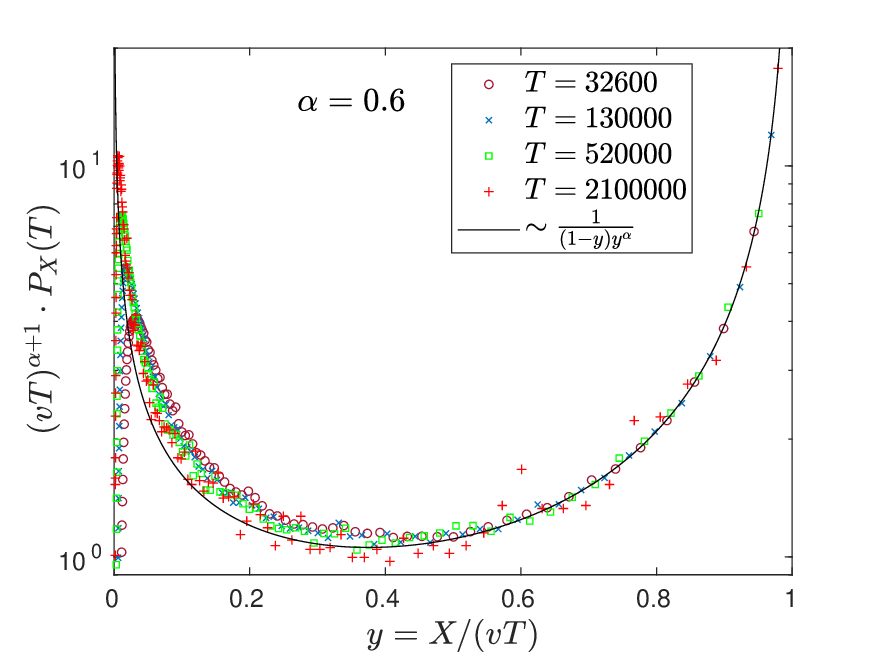}
\includegraphics[scale = 0.62]{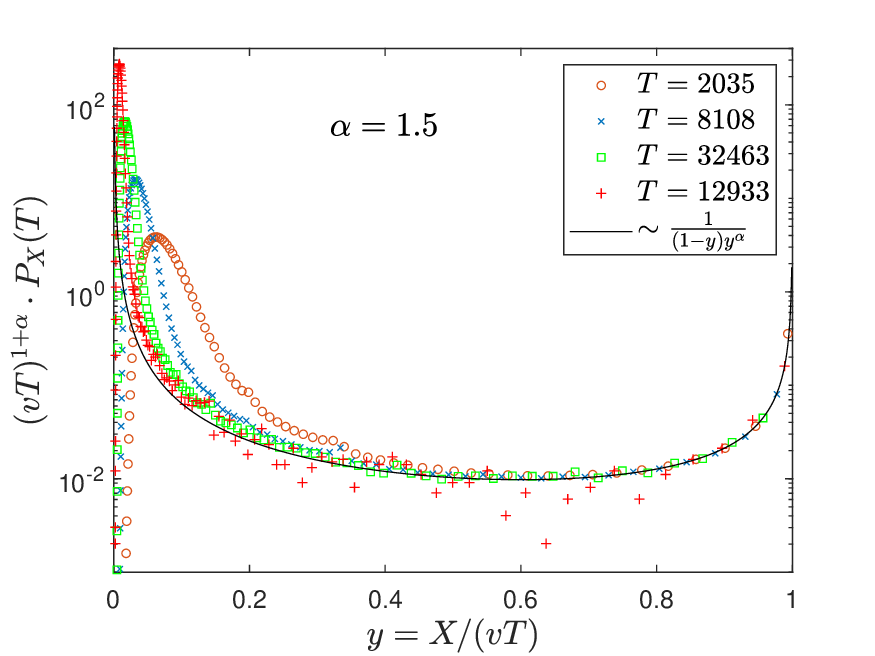}
\caption{Exit probability $P_X(n)$ for a L\'evy Lorentz gas. $P_X(n)$ is multiplied by $(vT)^{1+\alpha}$ and plotted as a function of the distance $X/(vT)$. Upper panel $\alpha=0.6$; lower panel $\alpha=1.5$. The continuous line shows the asymptotic analytical prediction in Eq. \eqref{eq:LevyLorentz}. In this case a multiplicative constant has been optimized to reproduce numerical data. {For large times the probabilities are small and the stochastic fluctuations become large, especially for $\alpha>1$.}}
\label{fig:LevyLorentz}
\end{figure}

{Here, $n(T_w)$ is the number of times (scattering events) in which a walker is reflected or transmitted by a scatterer. A renewal only occurs when a walker is transmitted for the first time by a scattering point, since in this case the distance of the next scatterer is a random variable, independent of the RW history. Then, the big jump principle \cite{BJ1} applies and the walker can overcome the distance $X\gg \ell(T)$ only by entering a region where the two scatterers are separated by a distance greater than $X$.} Since the motion is ballistic, the large jump occurs at $T_w=T-X/v$ and the rate of independent renewals is $R(T_w)\sim T_w^{-\frac{1}{1+ \tilde \alpha}}$ \cite{BJ1}. On the other hand, the Sparre-Andersen theorem gives $S(T_w)\sim n(T_w)^{-1/2}$. Without considering the lattice spacing, the dynamics is a RW on a lattice, and the number of different scatterers visited in $n(T_w)$ steps is $v(T_w)\sim n(T_w)^{1/2}$. Note that $v(T_w)$ is also the number of scatterers within a distance $\ell(T_w)$; so the result in \cite{LevyLorentzBeenakker} gives  
$v(T_w)
\sim \ell(T_w)^{\tilde{\alpha}}\sim T_w^{\frac{\tilde {\alpha}}{1+\tilde {\alpha}}}$ 
and $n(T_w)\sim T_w^{\frac{2 \tilde \alpha}{1+\tilde \alpha}}$
(for $\alpha\geq 1$ we recover that the number of scattering events is proportional to the elapsed time). 
Since two scatterers are separated by a distance greater than $X$ with probability $P_s(X)=\int_X^\infty dl p(l)$, Eq. \eqref{eq:BJ} gives:

\begin{equation}
    P_X(T)  \sim 
    \frac{1}{T_w^{\frac{ 1}{1+\tilde \alpha}}}\cdot
    \frac{1}{T_w^{\frac{ \tilde \alpha}{1+\tilde \alpha}}} \cdot
    \int_{X}^\infty dl p(l) 
    \sim \frac{1}{T^{\alpha+1}}
    \frac{1}{(1-y)y^\alpha}.
    \label{eq:LevyLorentz}
\end{equation}
where again the rescaled distance $y=X/(vT)\leq 1$ is induced by the ballistic motion in the single jump. 
Eq. \eqref{eq:LevyLorentz} contains a multiplicative constant since the rate and the number of jumps have been calculated only by scaling arguments. Figure 
\ref{fig:LevyLorentz} compares Eq. \eqref{eq:LevyLorentz} to numerical simulation, showing an excellent agreement.

{\it Conclusions}. 
{We formulate a big jump approach for first passage probabilities and use it to derive the analytical form of the tail in the exit distributions for three fundamental jump processes: the discrete time RW, the L\'evy walk and the L\'evy-Lorentz gas. In practice, the physical information about how this single large jump unfolds is used to give an exact analytical estimate by separating the dynamics at the jump from the whole stochastic process, which can also be very complex.} Our result shows that, in the presence of power law jump distributions, anomalous exit events can occur on time scales that are orders of magnitude smaller than typical exit times, even when the process is Gaussian. The result is heuristic and based on the application of the big jump principle, yet comparison with detailed numerical simulations suggests that the estimate is essentially correct, thus opening the way to a rigorous derivation.
{In this perspective, a fundamental question should be to identify what are the general characteristics of a stochastic process with subexponential distribution, allowing the application of the single big jump principle and in particular of Eq. \eqref{eq:BJ}. In its general form, the principle can also be applied to continuous-time processes and stochastic differential equation, provided that one can again identify the set of sub-exponential 'jumps', as has been successfully done in the case of the dynamics of cold atoms \cite{BJ1}. In that case, the big jump could be related to istantonic solutions \cite{Alqahtani,Naftali,Krajnick}.}  Finally, we obtained our result for power law distributions but we expect such estimates to hold for a large class of sub-exponential jump processes \cite{BJFoss,Goldie}, providing a boost to the study of rare events in first-passage probabilities.  

\begin{acknowledgments}
We warmly thank Olivier B\'enichou and J\'eremie Klinger for interesting discussions. 
\end{acknowledgments}




\providecommand{\noopsort}[1]{}\providecommand{\singleletter}[1]{#1}%

\end{document}